\documentclass[journal,comsoc]{IEEEtran}
\usepackage[utf8]{inputenc}
\usepackage{color}
\usepackage{acronym}
\usepackage{balance}
\usepackage[pdftex]{graphicx}
  \usepackage{epstopdf}
  \DeclareGraphicsExtensions{.pdf,.jpeg,.png,.eps}
\acrodef{sn}[SN]{satellite network}
\acrodef{uc}[UC]{use case}
\acrodef{an}[AN]{aerial network}
\acrodef{gn}[GN]{ground network}
\acrodef{rf}[RF]{radio frequency}
\acrodef{ris}[RIS]{reconfigurable intelligent surface}
\acrodef{vlc}[VLC]{visible light communications}
\acrodef{re}[RE]{reflective element}
\acrodef{geo}[GEO]{geostationary orbit}
\acrodef{meo}[MEO]{medium earth orbit}
\acrodef{leo}[LEO]{low earth orbit}
\acrodef{hap}[HAP]{high altitude platform}
\acrodef{lap}[LAP]{low altitude platform}
\acrodef{vn}[VN]{virtual network}
\acrodef{uav}[UAV]{unmanned aerial vehicle}
\acrodef{3d}[3D]{three-dimensional}
\acrodef{fso}[FSO]{free-space optical}
\acrodef{owc}[OWC]{optical Wireless Communications}
\acrodef{bs}[BS]{base station}
\acrodef{ber}[BER]{bit error rate}
\acrodef{thz}[THz]{terahertz}
\acrodef{mmwave}[mmWave]{millimeter wave}
\acrodef{ris}[RIS]{reconfigurable intelligent surface}
\acrodef{mimo}[MIMO]{multiple-input multiple-output}
\acrodef{los}[LoS]{line-of-sight}
\acrodef{nlos}[NLoS]{Non-LoS}
\acrodef{ir}[IR]{infrared}
\acrodef{uv}[UV]{ultraviolet}
\acrodef{ioe}[IoE]{Internet-of-Everything}
\acrodef{occ}[OCC]{optical camera communication}
\acrodef{lifi}[LiFi]{light fidelity}
\acrodef{led}[LED]{light-emitting diode}
\acrodef{ml}[ML]{machine learning}
\acrodef{qos}[QoS]{quality-of-service}
\acrodef{sagin}[SAGIN]{space-air-ground integrated network}
\acrodef{csi}[CSI]{channel state information}
\acrodef{em}[EM]{electromagnetic}
\acrodef{gan}[GAN]{generative adversarial network}
\acrodef{snr}[SNR]{signal-to-noise ratio}
\acrodef{sinr}[SINR]{signal-to-interference-plus-noise ratio}
\acrodef{e2e}[E2E]{end-to-end}
\acrodef{5g}[5G]{fifth generation}
\acrodef{mec}[MEC]{mobile edge computing}
\acrodef{mems}[MEMS]{microelectromechanical systems}
\acrodef{6g}[6G]{sixth generation}
\acrodef{4g}[4G]{fourth generation}
\acrodef{b5g}[B5G]{beyond 5G}
\acrodef{5gb}[5GB]{5G and beyond}
\acrodef{tbps}[Tbps]{Tera-bit-per-second}
\acrodef{ios}[IoS]{Internet-of-Senses}
\acrodef{xr}[XR]{extended reality}
\acrodef{bci}[BCI]{brain-computer interfaces}
\acrodef{urllc}[URLLC]{ultra-reliable low-latency communications}
\acrodef{iot}[IoT]{Internet-of-Things}
\acrodef{nems}[NEMS]{Nano Electro Mechanical Systems}
\acrodef{ict}[ICT]{information and communication technology}
\acrodef{un}[UN]{United Nations}
\acrodef{sdg}[SDGs]{sustainable development goals}
\acrodef{ai}[AI]{artificial intelligence}
\acrodef{agi}[AGI]{artificial general intelligence}
\acrodef{ran}[RAN]{radio access network}
\acrodef{dt}[DT]{digital twins}
\acrodef{kpi}[KPI]{key performance indicator}
\acrodef{mmtc}[mMTC]{massive machine-type communications} 
\acrodef{g}[G]{generation} 
\acrodef{nems}[NEMS]{nano electro mechanical systems }
\acrodef{sla}[SLA]{service level agreement}
\acrodef{p2p}[P2P]{point-to-point}
\acrodef{embb}[eMBB]{enhanced mobile broadband} 
\acrodef{urllc}[URLLC]{ultra-reliable low-latency communications} 
\acrodef{cds}[CDS]{compute, data, and storage}
\acrodef{sos}[SoS]{systems of systems}
\acrodef{cl}[CL]{continuous learning}
\acrodef{aci}[ACI]{ artificial collective intelligence}
\acrodef{iomust}[IoMusT]{Internet-of-Musical Things}
\acrodef{ng}[NG]{next-generation}
\begin{document}
\title{The Journey Towards 6G:\\A Digital and Societal Revolution in the Making} 


\author{Lina~Mohjazi,~Bassant~Selim,~Mallik~Tatipamula, and~Muhammad~Ali~Imran

\thanks{L. Mohjazi   and  M. A. Imran are with the James Watt School of Engineering, University of Glasgow, Glasgow, G12 8QQ, UK. (e-mail:\{Lina.Mohjazi, Muhammad.Imran\}@glasgow.ac.uk).}
\thanks{B. Selim is with École de Technologie Supérieure, Montreal, QC H3C 1K3, Canada (e-mail: bassant.selim@etsmtl.ca).}
\thanks{M. Tatipamula is with Ericsson Silicon Valley, Santa Clara, CA 95054, USA (e-mail: mallik.tatipamula@ericsson.com).}}

\maketitle

 \begin{abstract}
While the \ac{5g} is bringing an innovative fabric of breakthrough technologies, enabling smart factories, cities, and \ac{iot}, the unprecedented strain on communication networks put by these applications, in terms of highly cognitive, agile architectures and the support of massive connectivity, energy efficiency, and extreme ultra-low latency, is pushing \ac{5g} to their limits. As such, the focus of academic and industrial efforts has shifted toward \ac{b5g} and the conceptualization of \ac{6g} systems. This article discusses four main digital and societal \acp{uc} that will drive the need to reconcile a new breed of network requirements. Based on this, we provide our vision of the fundamental architectural ingredients that will enable the promise of \ac{6g} networks of bringing the unification of experiences across the digital, physical, and human worlds. We outline key disruptive technological paradigms that will support \ac{6g} materialize a bouquet of unique expectations and redefine how we live and protect our planet. Finally, we adopt the recently envisaged ecosystem of the \ac{iomust} to depict how the discussed \acp{uc} and technological paradigms may be exploited to realize this ecosystem.
\end{abstract}

\section{Introduction} 
The rollout of fifth generation (5G) wireless networks continues across the world 
and is bringing a technological breakthrough with respect to previous communication networks. Another substantial development is currently taking place through beyond 5G (B5G) systems, where the focus is shifting from rate-centric \ac{embb} services to \ac{urllc} and \ac{mmtc}. However, with the emerging revolutionary changes in individual and societal trends, it is arguable whether the capabilities of \ac{b5g} systems will keep up the pace with the market demands of 2030, which is expected to witness a new breed of \ac{ioe} services and radical advancements in human-machine interaction technologies \cite{edgenative}.\\
\indent Supporting cutting-edge future services not only necessitates delivering ultra-high reliability, extremely high data rates, and ultra-low latency, but also requires a new wave of network design, where computing, control, and communication are integrated into a network fabric that offers functionalities beyond communications, such as spatial and timing data. Researchers speculate that this grand vision will push \ac{5g} systems toward their limits within 10 years of their launch and calls for the conceptualization of sixth generation (6G) systems,  envisioned to introduce disruptive technologies and innovative network architectures that govern the evolution from connected everything to connected intelligence 
\cite{edgenative}. \\
\indent In this article, unlike existing works, which generally adopt a classic fashion in discussing \ac{6g} systems requirements, challenges, and enabling technologies, we outline the vision of a \ac{6g}-powered world by shedding light on what \ac{6g} promises in Sec.~\ref{promise} and overviewing the use cases (UC) that are expected to drive a digital and societal revolution in Sec.~\ref{usecases}. This is followed by introducing the paradigm shifts that formulate an evolved network architecture in Sec.~\ref{architecture}. In Sec.~\ref{tech}, we highlight the main \ac{6g} technologies needed to realize the vision of a future connected cyber-physical world. Finally, Sec.~\ref{IoMusT} focuses on the recently envisioned \ac{iomust} \cite{8476543} to depict how the discussed \acp{uc} and technological paradigms can realize this ecosystem.

\section{What Does \ac{6g} Promise?} \label {promise}

As the \ac{5g} is gaining ground, the research community has shifted towards laying the fundamentals of \ac{ng} systems. The \ac{6g} vision is to create a seamless reality where the physical and digital worlds, so far separated, are converged. This will enable seamless movement in a cyber-physical continuum of a connected physical world of senses, actions, and experiences, and its programmable digital representation. With this futuristic converged reality, new ways to work remotely, meet new people, and experience foreign cultures and places will be made possible \cite{ericsson6G}.

In addition, the vision of \ac{6g} is  to also create more human-friendly, sustainable, and efficient communities. This requires networks that guarantee worldwide digital inclusion to support a wide range of elements, \ac{e2e} life-cycle tracking to reduce waste and automate recycling, resource-efficient connected agriculture, universal access to digital healthcare, etc. This requires embedded autonomous sensors and actuators, worldwide coverage with outstanding energy-, material-, and cost-efficiency, as well as a network platform with high availability and security \cite{ericsson6G}.

\begin{figure*}[!t]
\centering
 \includegraphics[width=0.9\textwidth]{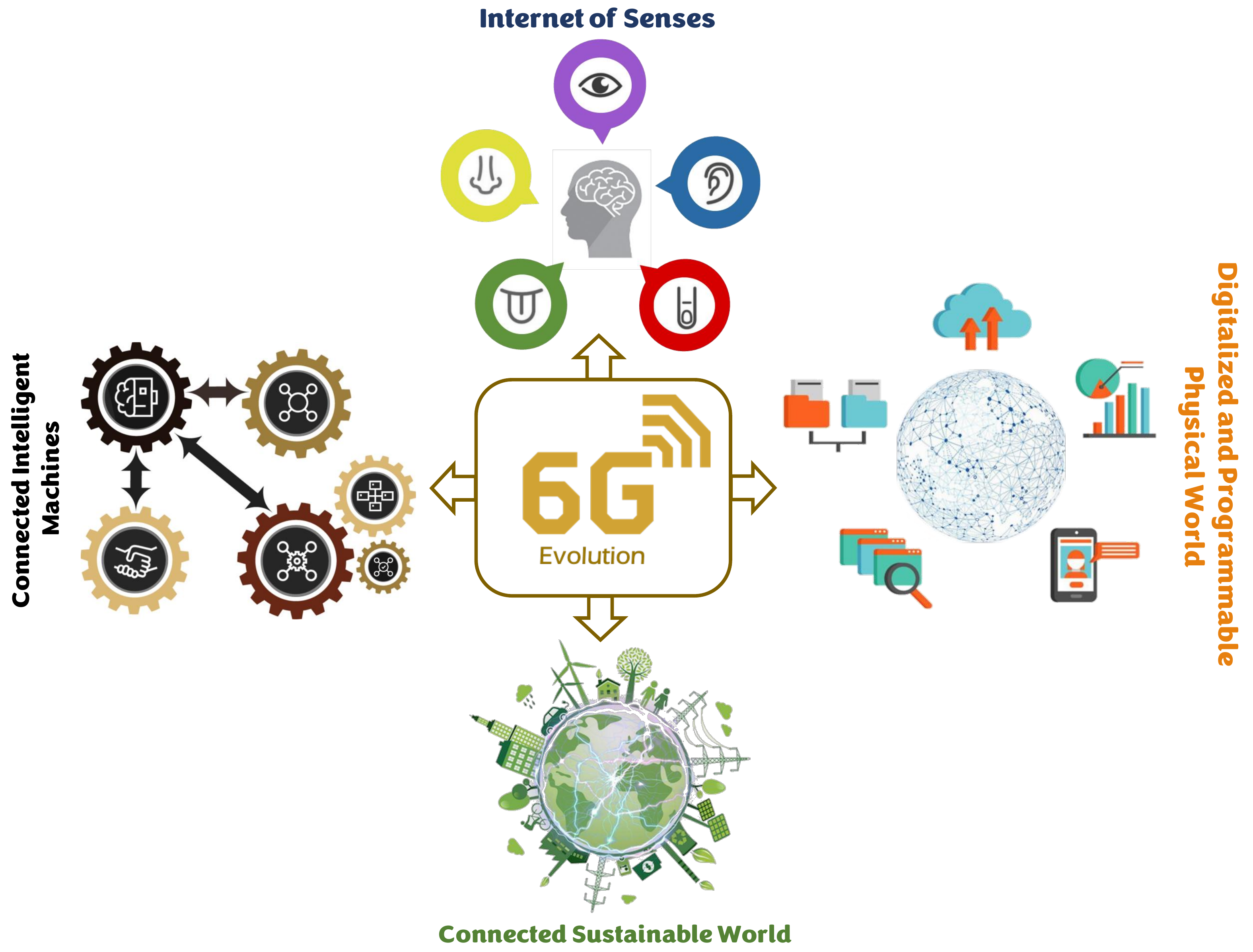}
 \caption{6G Use Cases.}
 \label{6G_Promise}
\end{figure*}

\section{Driving Trends and Use Cases (UCs)} \label{usecases}
The prospected \ac{6g} promise stems from four main driving \acp{uc}, which both lay the foundation for numerous applications and set the technical requirements for future communication systems. These \acp{uc} are shown in Fig.~\ref{6G_Promise} and are detailed in this section. 
\subsection{\ac{ios}: Any sense for anyone?}
The essence of the \ac{ios}, coined by Ericsson \cite{ios}, is built on delivering multisensory experiences over the networks. This \ac{uc} allows humans to have digital sensory immersive experiences that replicate or even augment what we experience in the physical world through visual, auditory, haptic, olfactory, and gustatory stimuli. This opens the door for an unprecedented opportunity to blend those multisensory experiences with our local surroundings and interact with people, devices, and robots remotely and in real time, without any location boundaries. The \ac{ios} is envisioned to be manifested through many revolutionary interactions, such as immersive communication, five senses merged reality, all senses online shopping, remote operation of machinery, Venn rooms, and sustainable vacations in virtual reality to name a few.

While \ac{ios} has the potential to enhance human interactions and thereby, bring people together, integrating telereality into the perceived world introduces non-trivial challenges, such as delivering perfect synchronization between all the sensory modalities and between the digital and real worlds, executing ubiquitous \ac{urllc}, and safeguarding users' privacy. As such, it is envisioned that the realization of \ac{ios} over the coming decade will be a result of the development of key stepping stones, namely, 1) \ac{ng} devices, sensing, and actuation, 2) multi-sensory and holographic communication, 3) network platform enablers for the \ac{ios}, 4) trustworthy \ac{ios}, and 5) \ac{bci} and shared biological experiences.

\subsection{Digitalized and Programmable Physical World}\label{digital} The tremendous growth in connected \ac{iot}, which is based on the real-time exchange of data generated from billions of embedded sensors and actuators, creates the fabric for the digital representation of the physical world. Leveraging this lays forward the foundation for having every imaginable object connected and digitally interactive in many activities and processes across industries and society at large. This also allows each physical action to be controlled and programmed according to detailed planning and execution of activities.  

In future, further digitalization will become possible through the emergence of new materials, such as metal oxides, polymers, etc., and sensor technologies, such as \ac{nems} enabling smart dust \cite{ilyas2018smart}. Moreover, wearable devices are expected to evolve towards more miniaturized form factors, to be integrated into skin tattoos and embedded in clothes, or printed onto different flexible bio-degradable substrates. Also, distributed ubiquitous sensing and actuation and associated computing and networking can be attached to real-world objects and environments or embedded inside objects. This will require intelligent data processing at the deep edge in single nodes or distributed across clusters of nodes. Exploiting advanced robotic systems will open new horizons for digitalization and programmability as they can be made available anywhere through the use of their onboard sensing, manipulation, locomotion and mechanical reconfiguration capabilities, with onboard computing, power and communication.

\subsection{Connected Intelligent Machines} 
The emergence of a new breed of data traffic is yet another aspect revolutionizing \ac{ng} networks. Specifically, so far, networks were designed to mainly serve people, and thus only interact with human intellect. Meanwhile, future mobile networks will be able to communicate with various intelligent entities, such as \ac{ai}-powered intelligent machines, and ultimately enable widespread intelligence, where humans and robots interact with each other to solve complicated tasks. These transitions will inevitably have impacts on the network design and requirements \cite{ericsson6G}.

A major stepping stone towards this goal is to enable machine intelligence, which requires researchers to step away from the conventional narrow \ac{ai} approach, which focuses on a specific task, towards \ac{agi}\ \cite{9390376}. With machines moving towards \ac{agi}, they will be able to learn how to perform multiple tasks and navigate, understand, and achieve their goals in unexplored environments. This will blur the line  between humans and intelligent machines. In addition, with \ac{aci}, machines will be able to control and interact with other machines. In this context, {optimized data formats and protocols} will be necessary to facilitate machine and human-machine collaborations in the machine intelligence and human-machine continuum.

\subsection{Connected Sustainable World}  
The focus on sustainable development is pushing governments and organizations to reconsider their regulations and operations to answer the populations' increasing concerns. In this context, the \ac{ict} sector has been in the spotlight both as an enabler for achieving the \ac{un} \ac{sdg} \cite{united2016sustainable} and as a substantial contributor to  global carbon emissions. Therefore, nowadays sustainable \ac{ict} efforts are broadly classified into sustainable \ac{ict} and \ac{ict} for sustainability. The former ensures that the \ac{ict} sector meets the sustainable development targets, and is concerned with the resource efficiency of the \ac{ict} sector as well as its impact on people, society, and the planet. Meanwhile, \ac{ict} for sustainability is centered around using the \ac{ict} sector as an enabler for sustainability, such as for inclusion, for the benefit of society, and for the environment. 

Despite \ac{5g}'s game-changing specifications, which achieved 10 times greater energy efficiency than \ac{4g} and enabled life-changing applications \cite{zhang20196g}, it is undeniable that the full potential is not reached, where global connectivity and carbon neutrality are not a reality yet. To this end, the connected sustainable world empowered by \ac{6g} will require 1) resource efficient \ac{ict}, through energy lean systems and network sharing; 2) responsible \ac{ict}, that ensures transparency and traceability, secures human rights, and guarantees children online protection; 3) \ac{ict} for inclusion, that provides equal digital opportunities and helps support education and digital literacy; 4) \ac{ict} for society, with resilient networks supporting the economic development and productivity; 5) \ac{ict} for the environment, facilitating a circular economy and protecting planetary boundaries. 
\begin{figure*}[!t]
\centering
 \includegraphics[width=1\textwidth]{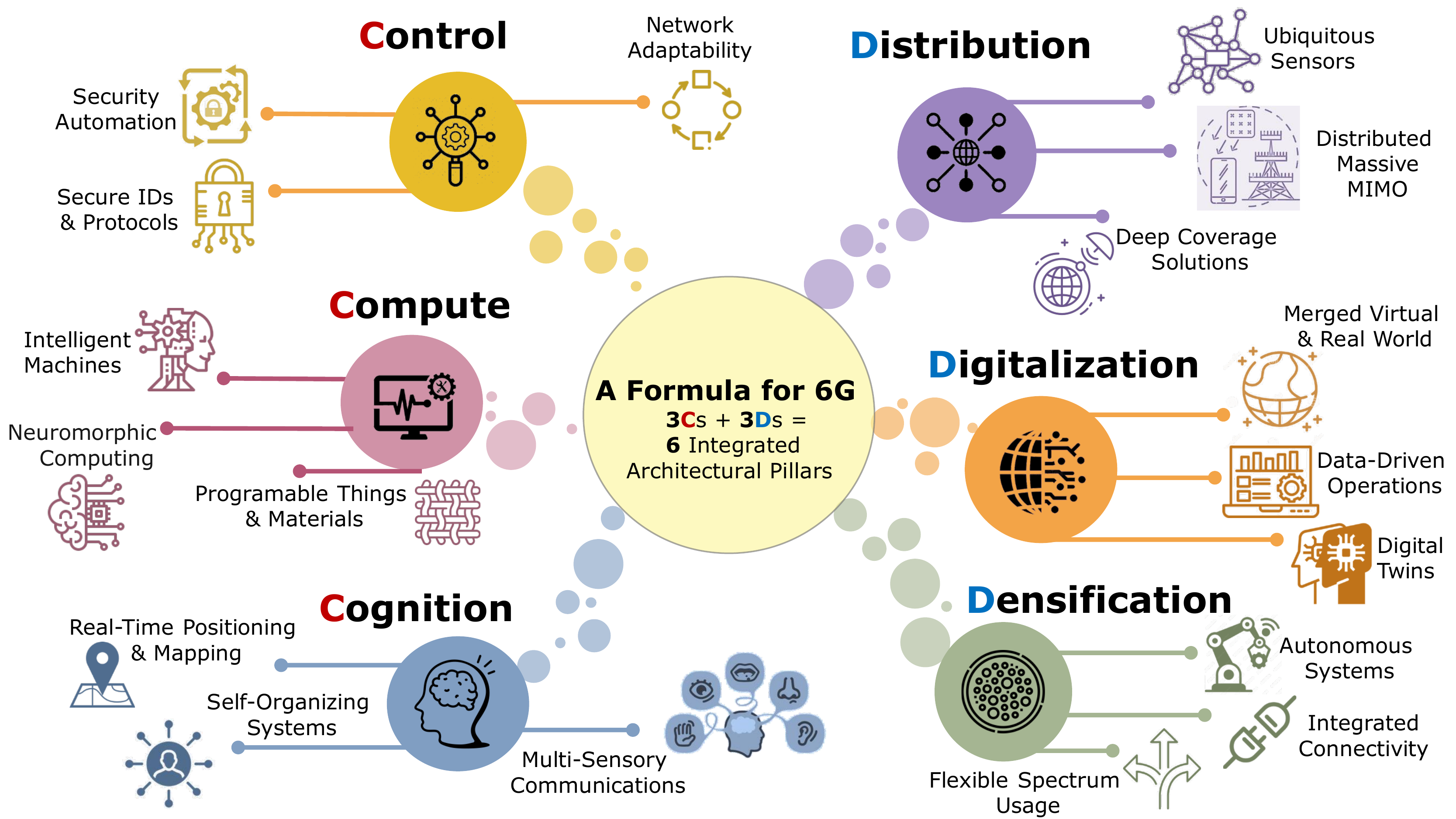}
 \caption{6G Architectural Pillars.}
 \label{6G_formula}
\end{figure*}
\section{Architecture Evolution: A Formula for 6G!} \label{architecture}

Capitalizing on the discussed  \acp{uc} and their requirements, in this section, we introduce six integrated architectural pillars for \ac{6g}, i.e. control, compute, cognition (3C) and densification, digitalization, distribution (3D). These pillars 
are presented in Fig.\ref{6G_formula}, and are discussed in details in what follows.

\subsection{Control}\label{control}
The complexity brought on by \ac{5g}, \ac{iot}, and the increase in the number of connected devices have pushed for virtualization of physical networks. New software-defined capabilities, including service assurance, orchestration, analytics, and data-centric processing, were brought on by this virtualization. This brings substantial benefits but also a great deal of complexity for managing and maintaining the network. In this context, it is clear that we need to step away from human-driven networks towards automation, which is essential for achieving streamlined services and operations, rock-solid reliability, and keeping businesses and users satisfied. Control systems are therefore now considered a pillar of \ac{b5g} networks, where special attention has been given to closed-loop automation \cite{9627832}. Closed-loop automation monitors and analyses network occurrences like failures and congestion using data and analytics, and then takes appropriate actions to resolve any issues. The term "loop" refers to the communication feedback loop between the network's performance being tracked, identified, adjusted, and optimized to allow for self-optimization.

Closed loops have been around and employed in control systems for a while. However, the idea of having many closed loops operating at once and interfering with one another is relatively new. This is particularly the case in autonomous communication networks, where this interference has to be identified and dealt with. In fact, numerous control loops are needed in order to optimize global networks and \ac{e2e} services. In this context, automatic conflict resolution, i.e. mechanisms to resolve conflicts between competing intents or closed loops autonomously are needed. As the network automation scale grows, certain control agents' goals will compete with one another, and there may be instances where several control loops are directed at the same controlled elements. In order to achieve the optimal trade-off between competing goals and achieve network and service management goals in \ac{b5g} networks, effective tools to resolve such conflicts are needed. 

\subsection{Compute}
As discussed in Sec.~\ref{digital}, it is expected that \ac{6g} \acp{uc} will widely rely on distributed ubiquitous sensing and actuation, which can be of diverse and new generations of modalities, leading the data processing to be highly distributed in tiers of different capabilities and include inferencing, \ac{ml}, sensor fusion, and sensing-actuation control loops to automate simple or richer tasks locally, all based on multi-modalities. A number of \acp{uc} will depend on \ac{bci}, meaning that, when interfacing directly with the brain, advanced algorithms will be needed to transform the signals from the brain to data that is meaningful to computers and vice versa. For example, enabling machines that sense and predict our future actions and desires requires advanced computation technologies empowered by prediction analysis, intent-based techniques of \ac{ml} (i.e. \ac{ml} that anticipate human intents), \ac{bci} that can access our thoughts, emotions, and memories. Additionally, \ac{6g} \acp{uc} are driven by a massive number of data and will, therefore, need massive computing power to infuse sensor data of real-world content into merged virtual reality. In dynamic environments, extremely low latency computation for dynamic object detection, tracking, and rendering is required to generate high-resolution spatial maps/moving object models.

It is envisioned that the network compute fabric will be distributed in nature and will extend all the way to even the tiniest devices, making these devices part of the fabric, termed "\textit{mist computing}" \cite{8960619}. \textit{Mist computing} is expected to be carried out at the extreme edge of the network in tiny devices, such as microcontrollers and sensors, and use computing power on  sensors to process, precondition, and optimize data before its transmission. This renders the processed data more compact and results in communication power savings. In this extended network \ac{cds} fabric, compute and storage always happen at the most optimal location in the entire system. Data and computing can also dynamically move or be distributed over different locations. The unified distributed \ac{cds} fabric created might lead to rethinking the architecture of the conventional \ac{iot} platform into a distributed fabric-native (or “mist-native”) digital world “platform“ consisting of tiny components distributed over the unified \ac{cds} fabric.

\subsection{Cognition} \label{congnition}

As both communications and radio sensing systems are evolving towards higher frequency bands, miniaturization, and larger antenna arrays, they are becoming increasingly similar in terms of signal processing, hardware architecture, and channel characteristics. This opens unprecedented opportunities for offering ubiquitous learning and intelligence across the network by integrating sensing within the existing wireless infrastructure. In particular, achieving a cognitive, self-learning physical world will rely on self-generating \ac{dt} of the planet, where almost everything can be programmed and repurposed, including physical materials \cite{9899718}. While this entitles the deployment of data-driven network architectures, where relevant data can efficiently and securely be distributed and analyzed to optimize the network performance and operations, it also means that decisions should be made on data whose exposure and collection are configured intelligently on-the-fly. As such, we need tools that automate the selection of source data for model training in natively distributed \ac{ml} algorithms. 

The \ac{ai} of the cognitive network might even be designing new services on the fly for machines based on their needs. The network could be seen as a large-scale distributed intelligent machine serving all other connected intelligent machines. From an architectural perspective, for the  network to be an enabler of cognitive systems, it would require a large-scale of distributed intelligent machines that provide connectivity, computing, storage, mediation, and other services to all other intelligent machines \cite{9899718}. It is, therefore, expected that ultimate cognition may be possible through the emergence of a connected global brain where the \ac{ai} of the network interacts with the \acp{ai} present in all those multi-agent \ac{sos} at a global scale, optimizing the operational aspects of the \ac{sos}. In this case, the network could also perform \ac{sos} governance and monitoring, including prediction, identification, and avoidance of unexpected emergent behaviors. 
This requires a shift from reactive to predictive management, which heavily depends on \ac{cl} for autonomous system operation, where  humans can take more of a supervisory role. Once this happens, a structure for governance should be introduced to decide what autonomy the management system is trusted with and how it should interact with its human operators.

\subsection{Densification}\label{densification}
One of the communication networks \acp{kpi} that has notably evolved over the past decades is connection density, which quantifies the number of devices satisfying a target \ac{qos} per unit area. For instance, \ac{5g} is expected to support 1,000,000 devices per square kilometer for \ac{mmtc} applications \cite{zhang20196g}, while 
 the connection density of \ac{6g} is expected to be at least an order of magnitude larger. Although not always in the spotlight, the connection density has significantly impacted the architecture of mobile networks. Indeed, increased connection density calls for higher spectrum efficiency, which has resulted in network densification through the deployment of increasingly smaller cells. 

{Network densification} adds more sites in a given area to increase the network capacity. This can be achieved through sector splitting or the deployment of  macro/small/indoor cells or by increasing the number of transmit/receive antennas. {Network densification} also lessens the effect of the high load on the hardware, which is not the case when spectrum modification is employed to satisfy the increasing connection density requirements. However, maximizing the efficiency, being the spectrum, energy, overhead, etc., of the network becomes increasingly complex with network densification. This is essential to enable cost-efficient deployment of very dense networks as well as to ensure the strict carbon neutrality targets discussed  earlier.

To this end, dynamic and flexible deployment solutions will become a cornerstone of \ac{6g} networks. This is essential for ensuring future-proof cost-efficient and sustainable high-capacity and resilient networks necessary for \ac{ng} \acp{uc}. In this context, heterogeneous nodes, including ad-hoc, user-deployed, mobile, and nonterrestrial ones, can be seamlessly integrated/removed to scale up/down the capacity. For instance, multi-hop deployment could enable cost-efficient on-demand or rural coverage.

\subsection{Digitalization}\label{Digitalization}

Ubiquitous digitalization and programmability will be automatically available anywhere and anytime by levering advanced distributed ubiquitous sensing and actuation. This enables a future where we can create digital replicas, or otherwise "\textit{\ac{dt}}" \cite{9899718} of everything we engage with in the real-world - like cities and their associated infrastructure and dynamics. \acp{dt} are expected to provide digital representations that may be accessible and controlled globally and simultaneously by a large number of individuals and applications, and will lay down the path toward detailed planning and execution of activities, as well as accurate management and optimization of large entities, such as buildings, roads, hospitals,  etc.    

Digital representation of everything is realizable through a number of cornerstones. More specifically, data structures and semantics of one physical entity should be harmonized and unified to allow for better utilization of the platforms to which \ac{dt} connect and to be able to extract more meaningful insights on the overall physical system. Moreover, \ac{dt} of all objects and places require evolved \ac{iot} platforms that make it possible to create \ac{dt} of even the simplest objects. Hence, having uniform data structure and ontology allows aggregation of twins into twins of larger systems, i.e. "\textit{massive twinning}", that facilitate semantic interoperability, where standards and \ac{ai} allow exchange between different systems and domains in a meaningful way \cite{9484096}. As such, systems that have not been jointly designed can extract the meaning of data from other systems. This is enabled by algorithms that can abstract meaning regardless of the information's representation format. 

Collaborative \ac{dt} of complex systems is a natural result of semantic interoperability, as the later enables \ac{dt} of different systems to engage in peer-to-peer
communications and jointly represent a larger portion of the physical reality. 
This in its turn opens the door for \textit{twin mobility} where twins of physical entities that change context or location follow their physical counterpart both in 1) where \ac{dt} reside, i.e., what compute and storage is used and 2) which \ac{dt} they interact with to model a bigger piece of reality, e.g., mobile user equipment connected to different networks or base stations, or a human seeking medical care in different locations
. 


\subsection{Distribution}

Recent advancements in virtualization of \ac{ran} and core network functions facilitated the management of highly distributed networks and led to the shift towards distributed architectures. This trend is accelerating with the increasing requirements for lower latency, higher volumes of users, and higher data rates, which are motivating the deployment of distributed network and compute elements closer to the users in \ac{5g}. In \ac{6g}, the distribution trend is expected to expand significantly both to push the performance, as in previous generations, and as a result of the technological advances, particularly the increased device processing capabilities and advances in \ac{ml}. Specifically, advances in sensing and actuation technologies, discussed in Sec. \ref{digital}, as well as the network densification and increased connection density presented in Sec. \ref{densification} will set the ground for futuristic highly distributed architectures. Meanwhile, new \ac{ml} and control paradigms, such as distributed learning  and \ac{aci}, will enable the  full potential of these distributed systems.    

In \ac{ng} networks, data storage and processing will therefore be highly distributed in layers with diverse processing capabilities, forming the network compute fabric. This architectural evolution will also affect the embedded intelligence in the applications and network. The traditional offline and centralized \ac{ml} is going to shift towards online models that are {natively} distributed in the different networked tiers for decentralized decision making. Such distributed architectures will also affect the way applications are processed in the network. This will require new methods for achieving seamless synchronization, mechanisms to support the \ac{e2e} performance guarantees of distributed applications, distributed data management and governance models, new efficient service-based interfaces, etc.

\section{A Technological Perspective} \label{tech}
In this section, we shed light on four key paradigms that are expected to shape the research focus on candidate innovative technologies that should be developed to facilitate the journey toward future network capabilities and \ac{ng} \acp{uc}' requirements.

\subsection{Limitless Connectivity}
 Emerging societal trends and applications relying on fully automated systems and intelligent services are validating the importance of telecommunication systems in delivering limitless connectivity  \cite{edgenative}. This encompasses both performance and coverage, meaning that connectivity should be available to anyone, anytime, and anywhere, and that the \ac{qos} should satisfy the application/user requirements in order to bridge the digital divide in a sustainable manner. As such, a fundamental enabling technology is network adaptability, which refers to the idea of enabling rapid network deployments and the fast introduction of new services. This comprises {dynamic network deployments}, which includes  ad-hoc or temporal deployments and mobile and non-terrestrial network nodes, enabling  cost-effective densification and limitless rural coverage.

In the context of densification and distribution, {lean network} design approaches aiming to reduce the overall network complexity, e.g.  by limiting the number of interfaces and the amount of duplicated functionality, are essential. Finally, the \ac{6g} architecture has to be {optimized for cloud deployments}. In fact, the \ac{5g} \ac{ran} architecture is based on nodes connected by point-to-point interfaces, while  the functional separation in the core was derived from pre-cloud legacy architectures \cite{9295415}. Therefore, both \ac{ran} and core service-based architectures will need to be re-engineered to include service-based interfaces that enable higher levels of optimization. Services supporting the dynamic and adaptable network vision will also be needed, including network sharing, network evolution, migration, etc.

\begin{figure*}[!t]
\centering
 \includegraphics[width=0.9\textwidth]{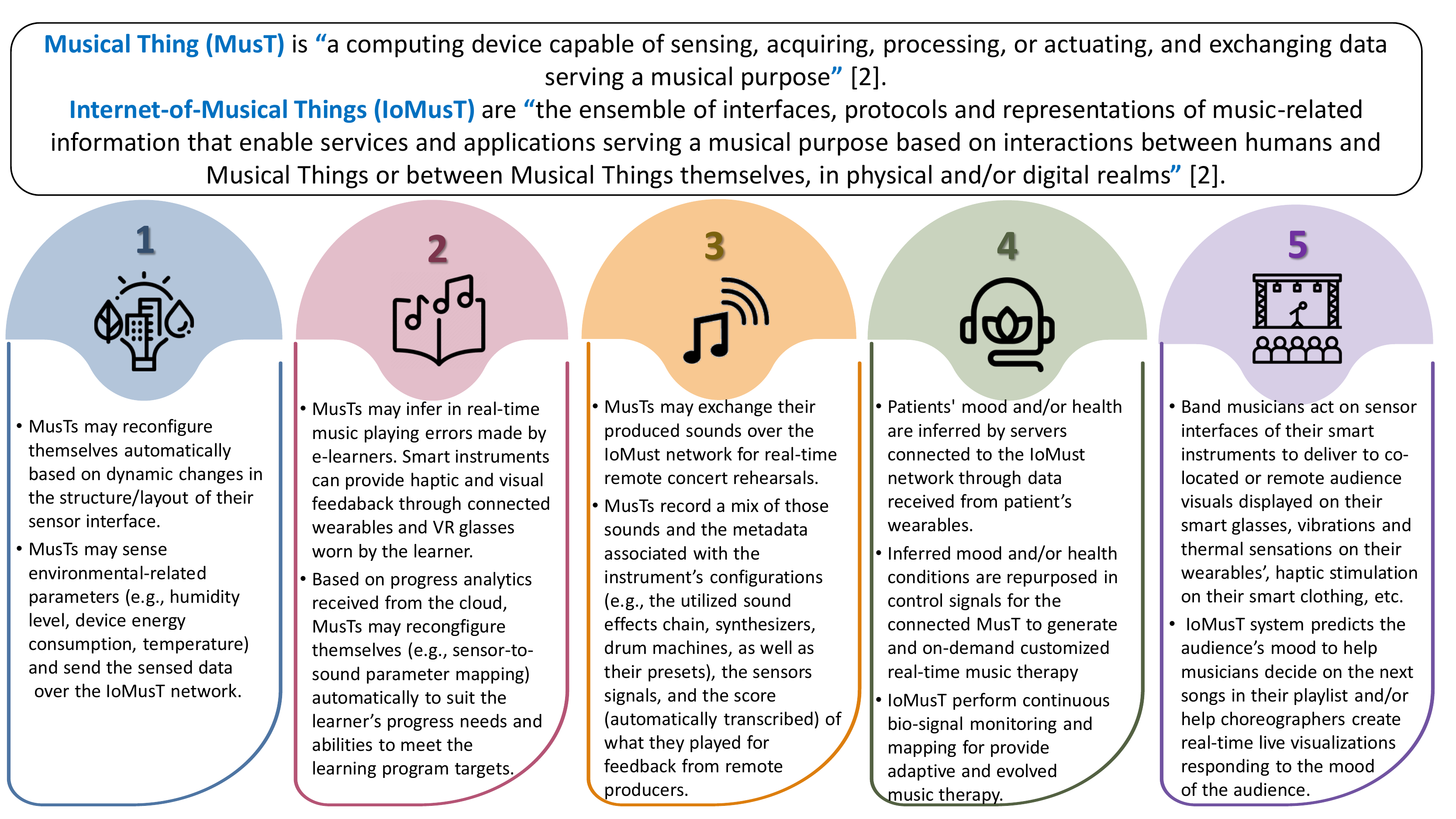}
 \caption{Internet-of-Musical Things overview and application scenarios: 1) Adaptation to Environmental Knowledge, 2) Enhanced Music E-Learning, 3) Remote Rehearsals, Intelligent Mixing, and Interaction with the Cloud, 4) Smart Music Therapy, and 5) Augmented and Immersive Live Concert Experiences \cite{8476543}.}
 \label{IoM1}
\end{figure*}
\begin{figure*}[!t]
\centering
 \includegraphics[width=0.9\textwidth]{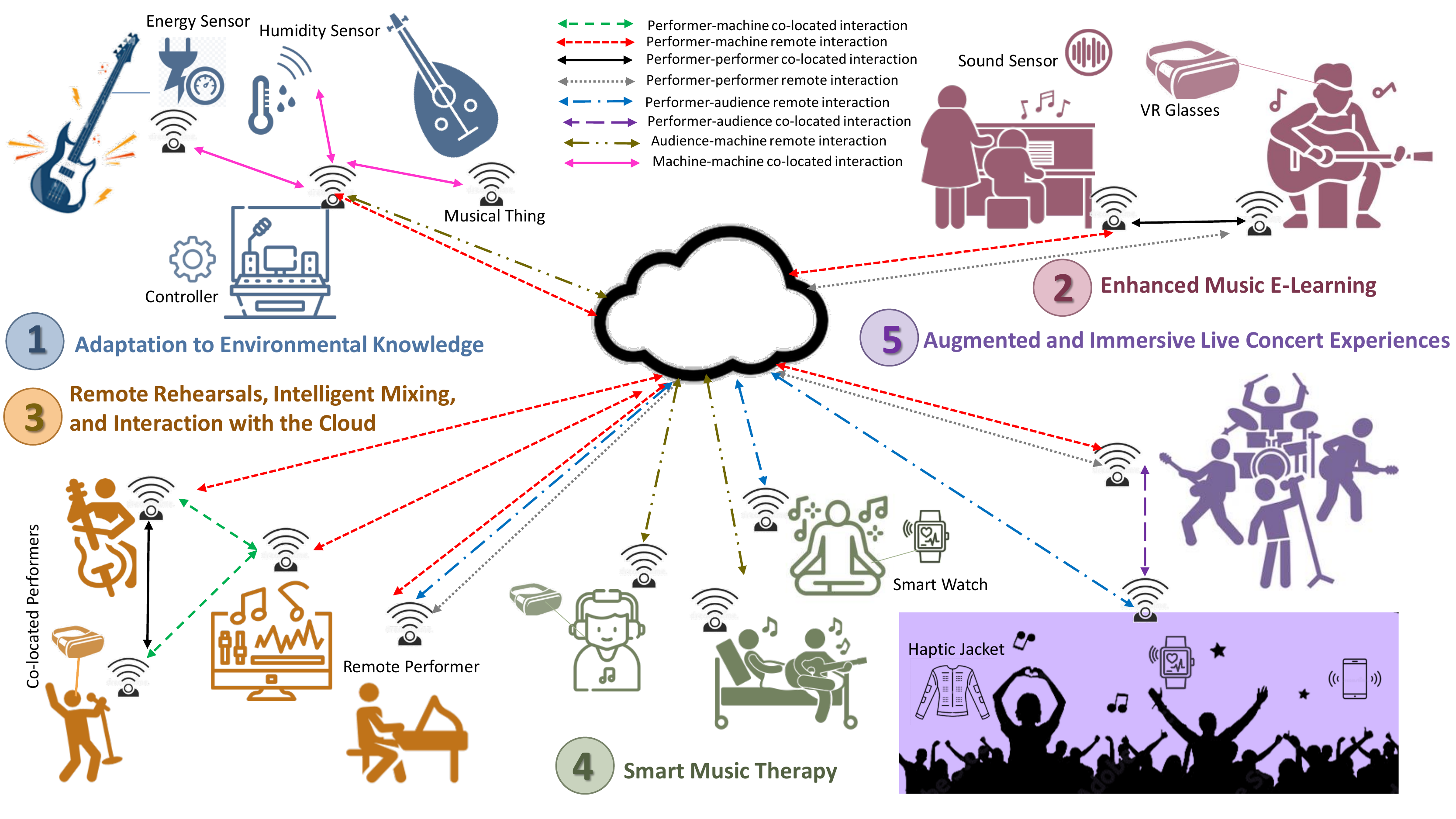}
 \caption{Possible interactions between musicians, audience members, and machines in the Internet-of-Musical Things ecosystem.}
 \label{IoM2}
\end{figure*}
\subsection{Trustworthy Systems}\label{trust}
Recently, we have witnessed mobile communications and networks shifting towards being a critical infrastructure supporting societies, industries, and consumers. As we move forward, network capabilities will evolve to the point where they can accommodate even the most mission- and business-critical \acp{uc}. This impels the need for {secure}, {resilient}, and {privacy-preserving} network platforms with proven satisfactory performance in challenging environments. All product lifecycle phases, including network development, deployment, and operation, must have built-in security automation and assurance capabilities \cite{ericsson6G}.

The level of confidence in the accuracy with which an information system's security features, practices, processes, and architecture effectively mediate and enforce the security policy is known as {security assurance}. 
In the evolution towards \ac{6g}, several challenges are expected to arise. The most notable one is certainly the security, privacy, and trustworthiness of \ac{ai}, which will be embedded in the network fabric. In fact, the security assurance of \ac{ai} and \ac{ml} is rather critical. Most often, \ac{ai} is regarded as a black box, which makes it hard to explain and complicates the security assurance, which for now can only be assessed by testing the output of a model for a given input. To safely embed \ac{ai} in critical applications and infrastructures, it is important to develop tools, processes, regulations, and benchmarks, to assess the security assurance of \ac{ai}. This includes assessing that the training and inference data is kept confidential and preserves privacy, the model does not reveal confidential or privacy-sensitive information through inversion or reverse engineering of trained models,  the model yields reliable and explainable outcomes, and the model and system integrity and robustness are maintained \cite{9524814}.

\subsection{Cognitive Networks}
To realize fully autonomous networks capable of supporting a wide spectrum of versatile services at no extra cost and complexity, it is crucial to raise the network intelligence level. This is expected to include two main aspects; 1) zero-touch deployment and operation and 2) continuous real-time performance improvements. This means that service and infrastructure configurations will no longer be handled manually.
 Instead, as discussed in Section \ref{congnition}, the network architecture will rely on distributed intelligence and \ac{cl} to make intelligent choices on \ac{e2e} service parameters. As discussed in Sec.~\ref{control}, new mechanisms need to be developed to manage the interaction of multiple closed loops running simultaneously and automatically resolve any conflicts resulting from this interaction.


Similarly, new technologies for intent-based management should be deployed to allow human-network interaction \cite{ericsson6G}. To this end, human operators use high-level declarative languages to define system operational goals in the form of "\textit{intents}, which are then translated to lower-level instructions (action plans and settings) using intelligence before they are applied in the network infrastructure. To realize this effectively, it is critical to develop cutting-edge reasoning, context awareness, and domain knowledge mechanisms capable of learning from the existing processed data and automatically making rational decisions and understanding how to apply an intent on different levels, e.g. domain, user, customer, application, operator, or country \cite{wang2021explainable}.   


\begin{figure*}[!t]
\centering
 \includegraphics[width=1\textwidth]{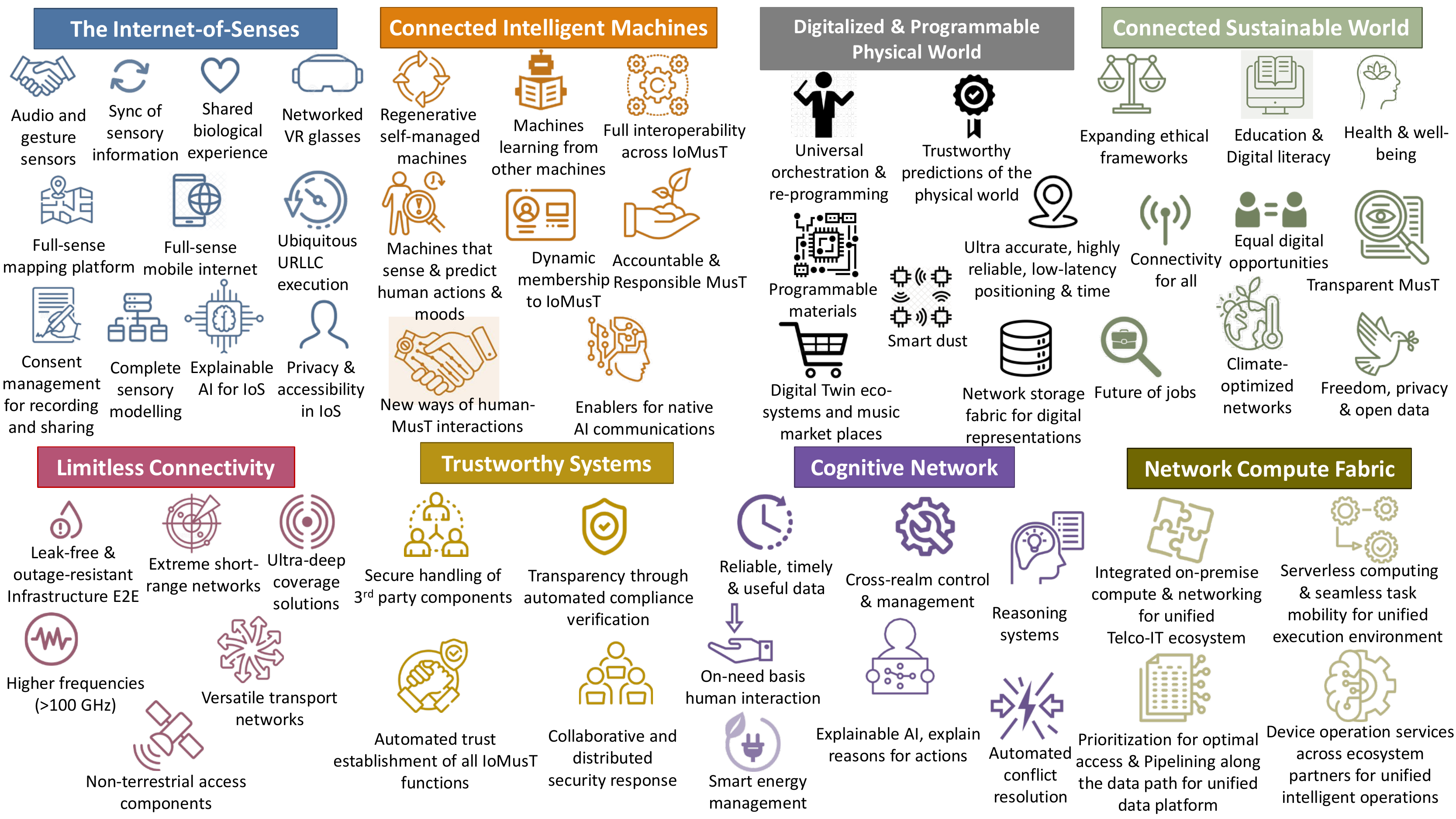}
 \caption{The Internet-of-Musical Things journey through 6G driving use cases and technological perspectives.}
 \label{IoMJourney}
\end{figure*}

\subsection{Network Compute Fabric}

The emerging \ac{6g} \acp{uc}, discussed in Sec.~ \ref{usecases}, require a pervasive network platform that tightly and effectively integrates the discussed six pillars formulating the \ac{6g} architecture. As such, for the network to act as one unified entity to satisfy the aforementioned pillars' requirements, innovative technologies should be developed to provision unification over ecosystems, execution environments, data platforms, and intelligent operations.

For the network compute fabric to be unified, it is anticipated that exposure and federation will happen across all the ecosystem partners including the network, application developers, cloud providers, operators, or device and equipment vendors \cite{ericsson6G,edgenative}. This means that exposure of integrated connectivity and computing to edge \acp{uc} and associated enablers will be required. Furthermore, standalone on-premise networks and edge cloud for digitalization of industry/enterprise are expected to be deployed in order to support real-time applications, keep data on-premise for security cost reasons, and guarantee resilience and scalability. Additionally, serverless computing paradigms will be needed to facilitate automation, where most of the development and deployment of distributed applications is performed on top of the network infrastructure. In this sense, regardless of the dynamic network changes, the application will always have access to a local computing service.

\section{A Spotlight on an Ecosystem: \ac{iomust}} \label{IoMusT}
To put the discussion presented in this article into the context of an ecosystem and inspired by the highly sought-after new era of entertainment experiences \cite{ericsson6G}, we illustrate the \ac{6g} driving \acp{uc} and technological perspectives for the recently envisioned futuristic \ac{iomust} ecosystem \cite{8476543}. An overview of \ac{iomust} and the relevant potential application scenarios are presented in Fig.~\ref{IoM1}. Furthermore, the associated possible interactions in this ecosystem are depicted in Fig.~\ref{IoM2}. It is not difficult to comprehend from Figs.~\ref{IoM1} and \ref{IoM2} that the deployment of \ac{iomust} paradigm will depend on a significant level of scalability in control, compute, cognition, distribution, digitalization, and densification to foster novel and diversified services settings, transparency and affordance, collaboration and communication, access and privacy, and a range of interaction types ranging from real-time interactive to highly asynchronous \cite{8476543}. Finally, Fig.~\ref{IoMJourney} shows several stepping stones, linking the \ac{iomust} journey with the \ac{6g} driving \acp{uc} and technological perspectives. 


\section{Conclusion} 

With \ac{5g} deployment progressing in leaps and bounds, the focus has shifted towards \ac{ng} communication networks. Although we are at a very early stage, where visions and predictions can be controversial, it is much anticipated that \ac{6g} will both build on the success points of previous generations and introduce revolutionary technologies and architectural shifts. This article presented a novel perspective on the \ac{6g} vision and the evolution of mobile networks towards this vision. To this end, we presented the \ac{6g} promise, portrayed through four driving \acp{uc}. To achieve the discussed promise, we introduced our vision of the six architectural pillars of \ac{6g} networks, and proceeded to highlighting key paradigms and corresponding technological advancements that will bring the anticipated \ac{6g} promise into reality.

\balance
\bibliographystyle{IEEEtran}

\bibliography{short}
\end{document}